\documentclass[aps,showpacs,amsmath,amssymb]{revtex4}
\usepackage[T2A]{fontenc}
\usepackage{amsmath}
\usepackage{amssymb}
\usepackage{graphicx}
\usepackage{euscript}

\begin{document}
\title{Chaotic mixing and transport in a meandering jet flow}

\author{S.V. Prants, M.V. Budyansky, M.Yu. Uleysky,} 
\affiliation{V.I.Il'ichev Pacific Oceanological Institute \\
of the Russian Academy of Sciences, 43 Baltiiskaya ulitsa, 690041 Vladivostok, Russia}
\author{and G.M. Zaslavsky}
\affiliation{Courant Institute of Mathematical Sciences, New York University, 251 Mercer St., 
New York, NY 10012, USA  \\
and Department of Physics, New York University, 2-4 Washington Place, New York, NY 10012, USA}
\begin{abstract}
Mixing and transport of passive particles are studied in a simple kinematic model 
of a meandering jet flow motivated by the problem of lateral mixing and transport in the Gulf 
Stream. We briefly discuss a model streamfunction, Hamiltonian advection equations, stationary points, 
and bifurcations. The phase portrait of the chosen model flow in the moving reference frame consists
of a central eastward jet, chains of  northern and southern circulations, and  peripheral 
westward currents. Under a periodic perturbation of the meander's amplitude, the 
topology of the phase space is complicated by the presence of chaotic layers and chains of oscillatory and ballistic islands with sticky boundaries immersed into a stochastic sea. 
Typical chaotic trajectories of advected particles 
are shown to demonstrate a complicated behavior with long flights in both the directions of motion 
intermittent with trapping in the circulation cells being stuck to the boundaries of  vortex cores and  
resonant islands. Transport is asymmetric in the sense that mixing between the circulations and the peripheral currents is, in general, different from mixing between the circulations and the jet. The transport properties 
are characterized by probability distribution functions (PDFs) 
of durations   and lengths of flights. 
Both the PDFs exhibit at their tails power-law decay with 
different values of exponents.  
\end{abstract}
\maketitle

\section{Introduction}
{\bf Passive particle advection can be considered as a Hamiltonian dynamical 
problem. Advection in a two-dimensional meandering shear flow, a simple kinematic model 
of lateral mixing and transport in the Gulf 
Stream, is an example of  a system with 3/2 degrees of freedom. The 
phase space of an advected particle in such a flow is mixed, i.e. it 
consists of chaotic layers, islands of periodic motion, and layers of 
periodic open trajectories. All these parts of the phase space influence 
  the transport of passive particles, which can become anomalous if 
trajectories stick to the boundaries of topological structures with 
regular dynamics. Particularly the stickiness to the so-called ballistic 
islands can create flights along the stream flow, make the transport 
superdiffusive, and influence in a strong way the advected particles 
distribution.}

\centerline{\rule{0.4\textwidth}{1pt}}

The goal of this paper is to study Lagrangian lateral
mixing and transport of passive particles
advected by a shear flow that has a coherent space
organization with two chains of vortices and
a strong jet between them. Temporal perturbations
of shear flows are known to induce chaotic mixing. Similar flows
occur naturally in the ocean as western boundary currents,
like the Gulf Stream in the Atlantic Ocean and
the Kuroshio in the Pacific Ocean, separating
water masses with different physical and biogeochemical
characteristics.

Hamiltonian chaos theory has been widely used to study advection
of passive particles in hydrodynamic flows of ideal fluid
(for review see \cite{Ar02,Ot89}). The equations of motion of a passive
particle advected by a two dimensional incompressible
flow is known to have a Hamiltonian form
\begin{equation}
\begin{aligned}
\frac{dx}{dt}=u(x,y,t)&=-\frac{\partial\Psi}{\partial y},\\
\frac{dy}{dt}=v(x,y,t)&=\frac{\partial\Psi}{\partial x},
\end{aligned}
\label{adv_eq}
\end{equation}
with the stream function $\Psi$ playing the role of a Hamiltonian
and the coordinates $(x,y)$ of a particle playing the role of canonically
conjugated variables. Equations (\ref{adv_eq}) form a Hamiltonian dynamical system 
whose phase space is identified as the physical space of advected particles. 

All time-independent one-degree-of-freedom  Hamiltonian systems are known to be 
integrable. It means that all fluid particles move along streamlines of a
time-independent streamfunction in a regular way. Equations ~(\ref{adv_eq}) with a time-periodic
streamfunction are usually non-integrable, giving rise to
chaotic particle's trajectories. In other words, even regular Eulerian velocity
fields $(u,v)$ can generate chaotic trajectories, the
phenomenon known  under the names ``chaotic advection'' or ``Lagrangian chaos'' 
\cite{Ar84,Ar02}.
In this field mixing means stretching and deformation of material
lines by advection. The term ``transport'' refers
to the motion of passive particles from one region of the physical space to
another.

Well developed theory of Hamiltonian chaotic mixing and transport
in the phase space (for a review see \cite{AKN88,MM87,Z02,Z05})
has been successfully applied to study chaotic
mixing and transport of passive particles in fluids
\cite{RLW90,WK89,Pi91,CFPV91,ZSW93,CNM93,Ko00,BB01,BJ98,BUP02,BUP04,LA05,BCL06}.
The phase space of a typical chaotic Hamiltonian system consists
of regions of chaotic mixing and regions
with regular trajectories. A boundary between them often contains a complicated
hierarchical structure of islands with regular trajectories
and can be extremely sticky for some values of control parameters
\cite{K83,CS84,M92,BZ93,BKWZ97,Panoiu}. All these properties are manifested
in mixing and transport of passive particles in fluids and have been 
observed in laboratory \cite{SMS89,SWS93,SWS94}. Particles
in the core regions of regular motion and in the islands are
trapped there forever unless we do not take into account molecular
diffusion and noise. They do not participate in the transport.
So the outermost KAM tori play the role of transport barriers in
the fluid. The particles, entering the boundary layer, can spend a very
large time there giving rise anomalous transport
or anomalous diffusion.

The motion in chaotic regions is extremely sensitive to small
variations in initial particle's positions. So one forces to use an statistical
approach to describe transport. A commonly
used statistical measure of transport is
the variance $\sigma^2(t)=\left<x^2\right>$, which for simplicity
is written for the particle displacement in one dimension.
The averaging is supposed to be done over an ensemble of
particles. 
If mixing in the phase space would be ideal (as it occurs in a hyperbolic
dynamical system with unstable orbits only, whose phase space does not contain 
any regular islands) the variance would satisfy the law of
normal diffusion, $\sigma^2(t)=2Dt$, with a well-defined diffusion
coefficient $D=\lim\limits_{t\to \infty}{\sigma^2/2t}$.
If trajectories are dominated by sticking 
to the boundaries of regular regions, where particles may spend
a long time, subdiffusion takes place with the transport law
$\sigma^2(t)\sim t^\mu$, $\mu<1$.
Superdiffusion with $\mu>1$ occurs when particles execute long ``flights'', i.e. travel
long distances between (or even without) sticking events.
Anomalous diffusion implies the diffusion coefficient
$D$ to be either zero ($\mu<1$) or infinite ($\mu>1$). As to transport 
of advected particles in fluids with diffusion \cite{BCVV95} and transport in 
the phase space in the presence of noise \cite{KRW82,MW95}, 
the situation is more compicated. For example, as it is shown in Ref. 
\cite{BCVV95} subdiffusiion is not possible  in incompressible flows 
in the presence of molecular diffusitivity.

It is as well possible to describe transport in terms of spatial extensions and durations 
of sticking and/or flight events. The respective 
probability density functions (PDFs) of durations of flights, $P(t) \sim t^{-\gamma}$, and of  
lengths of flights, $P(x) \sim x^{-\nu}$, are expected to be power-law functions. Long-time flights 
provide a link between chaotic advection and anomalous diffusion. Quantitative connections between 
all the exponents $\mu$, $\gamma$, and $\nu$ have been established with some simple models of 
chaotic motion and continuous-time random walks (for a review of anomalous transport  see \cite{MS84,SZK93,MK00,Z02}).

To specify the model we choose the shear flow in the form of the
Bickley jet \cite{B37} with the velocity profile $\operatorname{sech}^2 y$ and
an imposed running wave with the periodically
varied amplitude. This model is related to the so-called non-twist Hamiltonians 
(see \cite{CNM93,HH95,Simo} and references therein) that have a specific phase 
space topology and bifurcations with respect to the parameters change. 
As it usually occurs in Hamiltonian systems, separatrices
are destroyed in the presence of an arbitrary small time-dependent 
perturbation. Instead of them narrow stochastic layers appear which 
are ``seeds'' of Hamiltonian chaos. There are an infinite number of 
nonlinear resonances between the frequency of the perturbation and 
frequencies of rotations of advected particles. With increasing the 
strength of the perturbation, more and more of them begin to destroy 
and overlap. Motion of particles in the regions of the overlapping resonances 
in  the phase space is no more confined by impenetrable barriers 
of KAM tori and a large-scale transport becomes possible. 

In Sec.~\ref{Model} we
derive equations of motion for passive particles in the moving frame of
reference, find stationary points of the flow, topologically
different regimes of motion, and bifurcations between them.
Similar kinematic models with imposed time-periodic velocity 
fields have been used in literature on physical oceanography to
study lateral Lagrangian mixing
and transport in the meandering Gulf Stream current
\cite{B89,S92,DP94}. Mixing has been
shown with the help of Poincar{\'e} sections to be chaotic
in a wide range of the control parameters, and the
magnitudes of fluxes have been estimated with
the help of the Melnikov integral.
                   
In Sec.~\ref{Topology} we describe topology
of the phase space with regions of regular and chaotic
motion of advected particles. Sticking to the
boundaries of different types of islands, including ballistic ones,
and of the vortex cores is demonstrated on Poincar{\'e} sections of the
particle's trajectories.
Sec.~\ref{Transport} is devoted to quantifying
chaotic transport. We compute the statistics of durations and lengths of flights 
for a number of long-time trajectories and find the transport both in western and eastern 
directions to be anomalous with different values of the exponents.

\section{\label{Model}Model streamfunction, stationary points, 
and bifurcations}

We take the Bickley jet with a running wave imposed
as a model of a meandering shear flow in the ocean
\cite{B89}. The respective streamfunction in the laboratory
frame of reference has the following form:
\begin{equation}
%\begin{multline}
\Psi'(x',y',t')=-\Psi'_0\tanh{\left(\frac{y'-a\cos{k(x'-c t')}}
{\lambda\sqrt{1+k^2 a^2\sin^2{k(x'-c t')}} } \right)},
%\label{psi_prime}
%\end{multline}
\label{psi_0}
\end{equation}

where the hyperbolic tangent produces the velocity profile
$\sim\operatorname{sech}^2 y'$, the square root provides a possibility to have
a homogeneous Bickley jet with the width $\lambda$, and $a$, $k$, and $c$
are amplitude, wavenumber, and phase velocity of the wave,
respectively. The normalized streamfunction in the frame moving with the
phase velocity $c$ is 
\begin{equation}
\Psi=-\tanh{\left(\frac{y-A\cos x}{L\sqrt{1+A^2\sin^2 x}}\right)}+Cy,
\label{psi}
\end{equation}
where $x=k(x'-ct')$, $y=ky'$, $A=ak$, $L=\lambda k$, and
$C=c/\Psi'_0 k$.
\begin{figure}[!htb] 
\centerline{\includegraphics[width=0.8\textwidth,clip]{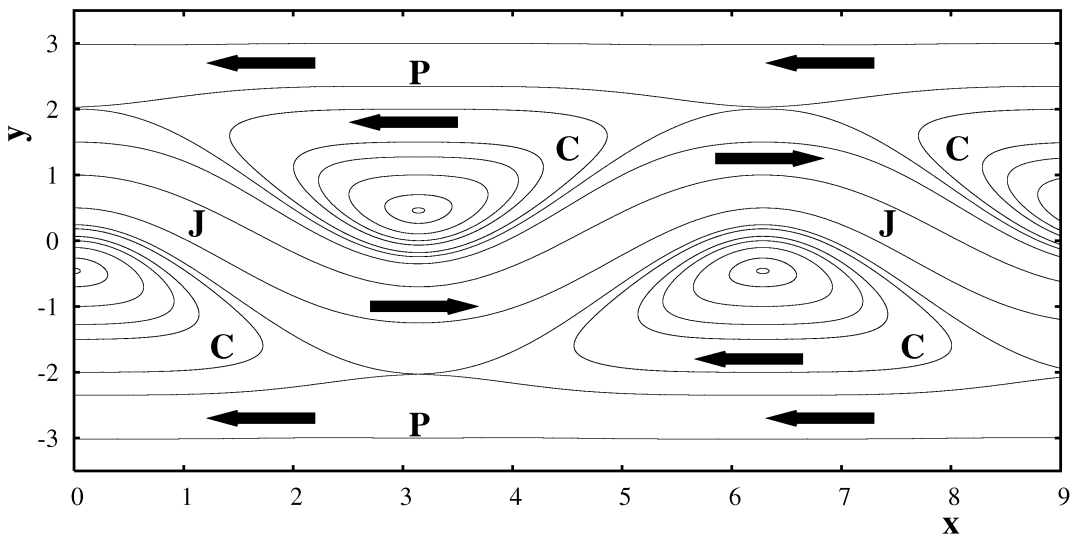}} 
\caption{ 
The phase portrait of the model shear flow (\ref{psi}) in the frame
moving with the meander's phase velocity.
Streamlines in the circulation ($C$),
jet ($J$), and peripheral currents ($P$) zones are shown.
The parameters are $A=0.785$, $C=0.1168$, and $L=0.628$.
} 
\label{fig1} 
\end{figure} 

The respective advection equations (\ref{adv_eq}) in this moving frame
have the form
\begin{equation}
\begin{gathered}
\begin{aligned}
\dot x&=\frac{1}{L\sqrt{1+A^2\sin^2 x}\cosh^2\theta}-C,\\
\dot y&=-\frac{A\sin x(1+A^2-Ay\cos x)}{L\left(1+A^2\sin^2 x\right)^{3/2}
\cosh^2\theta},
\end{aligned}\\
\theta=\frac{y-A\cos x}{L\sqrt{1+A^2\sin^2 x}},
\end{gathered}
\label{main_sys}
\end{equation}
where dot denotes differentiation with respect to dimensionless time
$t=\Psi'_0 k^2 t'$. The one-degree-of-freedom
dynamical system (\ref{main_sys}) is generated by the Hamiltonian 
(\ref{psi}) and has three control parameters~--- the jet's width $L$,
meander's amplitude $A$, and phase velocity $C$. The scaling
chosen results in the translational invariance of
equations (\ref{main_sys}) along the $x$-axis with the period $2\pi$.

A brief description of stability of stationary points and bifurcations of equations   
(\ref{main_sys}) is given below. It is clear from the first equation in
(\ref{main_sys}) that stationary points may exist only if the condition
$LC\le 1$ is fulfilled. Two equalities, following from the second
equation in (\ref{main_sys})
\begin{equation}
\sin x=0, \qquad 1+A^2-Ay\cos x=0, 
\label{cond1}
\end{equation}
give four stationary points
\begin{equation}
\begin{gathered}
x_{1,2}=0,\qquad   y_{1,2}=\pm L\operatorname{Arcosh}\sqrt{\frac{1}{LC}}+A,\\
x_{3,4}=\pi,\qquad y_{3,4}=\pm L\operatorname{Arcosh}\sqrt{\frac{1}{LC}}-A.
\end{gathered}
\label{points}
\end{equation}
As follows from the stability analysis, the second and third points are always
stable whereas the first and fourth ones are stable only if
\begin{equation}
AL\operatorname{Arcosh}\sqrt{\frac{1}{LC}}>1.
\label{cond2}
\end{equation}
It is proved in Ref.~\cite{UBP06} that if the condition
\begin{equation}
C<\frac{1}{L\cosh^2(1/AL)}
\label{cond3}
\end{equation}
is fulfilled, then there are four new stationary points in equations
(\ref{main_sys}) in addition to the points (\ref{points}).  

Having found the stationary points, their stability
properties, and how they are connected, we get the following topologically
different phase portraits of the streamfunction (\ref{psi})  
\begin{enumerate}
\item $C>C_\text{cr1}=1/L$, there are no stationary points.
\item[2a.] $C_\text{cr1}>C>C_\text{cr2}=1/L\cosh^2(1/AL)$ and 
$C>C_\text{cr3}$, there are four stationary points (\ref{points}):
two centers (the second and third ones) and two saddles
(the first and fourth ones). There are two separatrices each of which  
passes through its own saddle. 
A free flow between the separatrices is westward. 
\item[2b.] $C_\text{cr1}>C>C_\text{cr2}$ and $C<C_\text{cr3}$,
the stationary points are the same as in the preceding case, but a free
flow between the respective separatrices is eastward.
\item[3a.] $C_\text{cr2}>C>C_\text{cr3}$, there are eight stationary points:
four centers (\ref{points}) and four saddles with positions found in
Ref.~\cite{UBP06}. There are two separatrices, each of which connects two
saddle points. A free flow between the separatrices is westward.
\item[3b.] $C_\text{cr2}>C$ and $C<C_\text{cr3}$,
the stationary points are the same as in the preceding case, but a free
flow between the respective separatrices is eastward.
\end{enumerate}

Therefore, from the point of view of existence and stability of stationary
states, there are three possibilities:
1) there are no stationary points; 2) there are four stationary points,
two centers and two saddles; 3) there are eight stationary points, four
centers and four saddles. A bifurcation between the first and second regimes 
consists in arising two pairs ``saddle-center''. A bifurcation between 
the second and third regimes consists in arising two saddles and
a center between them instead of one saddle (a fork-type bifurcation). 

There is one more bifurcation that does not change the number and stability of stationary points but changes the topology of the flow. The values of the streamfunction 
(\ref{psi}) on the separatrices are equal on modulo but of opposite signs. There is a critical value of the phase velocity $C=C_\text{cr3}$ under which the separatrices  coincide 
and the respective streamfunction is equal to zero. If $C>C=C_\text{cr3}$, a free flow between the separatrices is westward, whereas with $C<C=C_\text{cr3}$ it is eastward.  It is difficult to find $C=C_\text{cr3}$ analytically but it may be shown \cite{UBP06} that $C_\text{cr3}>C_\text{cr2}$, if 
\begin{equation}
\frac{2(1+A^2)}{AL\sinh(2/AL)}<1.
\label{cond_Ccr3}
\end{equation}
Otherwise $C_\text{cr3}<C_\text{cr2}$. 
\begin{figure}[!htb] 
\centerline{\includegraphics[width=0.8\textwidth,clip]{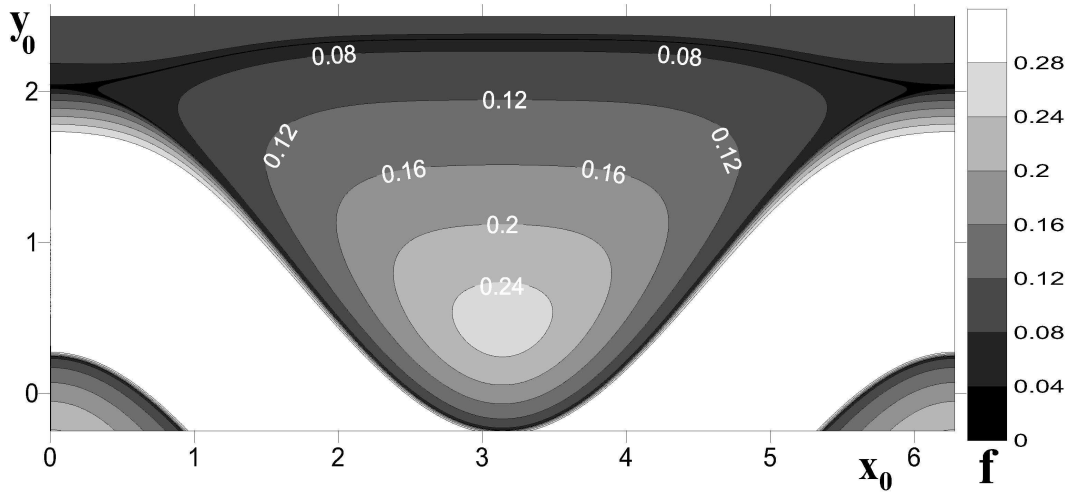}} 
\caption{ 
The frequency map represents by color the values
of frequency of rotation $f$ of particles with
initial positions ($x_0, y_0$) advected by the unperturbed flow (\ref{psi}).
} 
\label{fig2} 
\end{figure} 
\section{\label{Topology}Topology of the phase space}

Being motivated by the boundary oceanic eastward meandering jet currents
like the Gulf Stream and the Kuroshio, we will deal further in this
paper with the case 2b in the list in the preceding section.
The respective phase portrait in the reference frame, moving with the phase
velocity of the meander (Fig.~\ref{fig1}), consists
of three different regions: the central eastward jet ($J$), chains
of the northern and southern circulations ($C$), and the peripheral 
westward currents ($P$). The elliptic points (centers) of the circulations
are at critical lines $y=y_c$ with
$u(y_c)=c$ and $v(y_c)=0$. The northern separatrix connects
the saddle points at $x_1=2\pi k$ and $y_1$ to be defined
from equation (\ref{points}) and the southern 
one connects the saddle points at $x_4=(2k+1)\pi$ and $y_4$ from
(\ref{points}), where $k=0,\pm1,\dots$.
\begin{figure}[!htb] 
\centerline{\includegraphics[width=0.8\textwidth,clip]{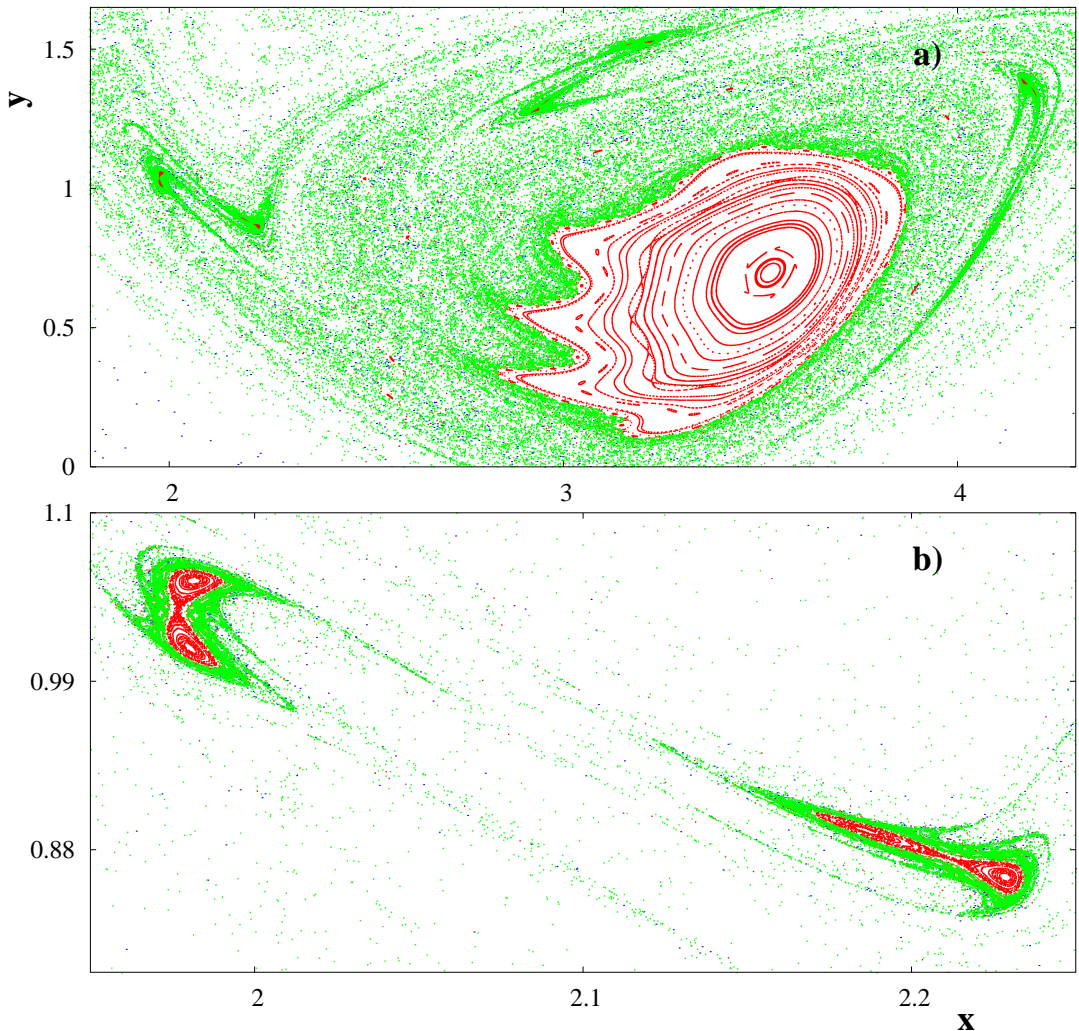}} 
\caption{ (color online). 
Poincar{\'e} section of the streamfunction (\ref{psi})
with the same parameters as in Fig.~\ref{fig1} and the amplitude
$\varepsilon=0.0785$ and frequency of perturbation $w=0.2536.$ 
(a) General view, (b) zoom of a pair of islands of the secondary resonance.
} 
\label{fig3} 
\end{figure} 

In the presence of a time dependent perturbation, the separatrices
are destroyed and transformed into stochastic layers. As a
perturbation, we take the periodic modulation of the meander's
amplitude, $A=A_0+\varepsilon\cos(\omega t+\varphi)$, in
the streamfunction (\ref{psi}) written in the frame moving with the phase
velocity of the meander. The jet's width $L=0.628$, the phase velocity $C=0.1168$, 
and the meander's amplitude $A_0=0.785$
will be fixed in simulation throughout the paper. These values are in the range of the parameters 
which has been estimated \cite{B89,S92} to be realistic with the Gulf Stream current. 
\begin{figure}[!htb] 
\centerline{\includegraphics[width=0.8\textwidth,clip]{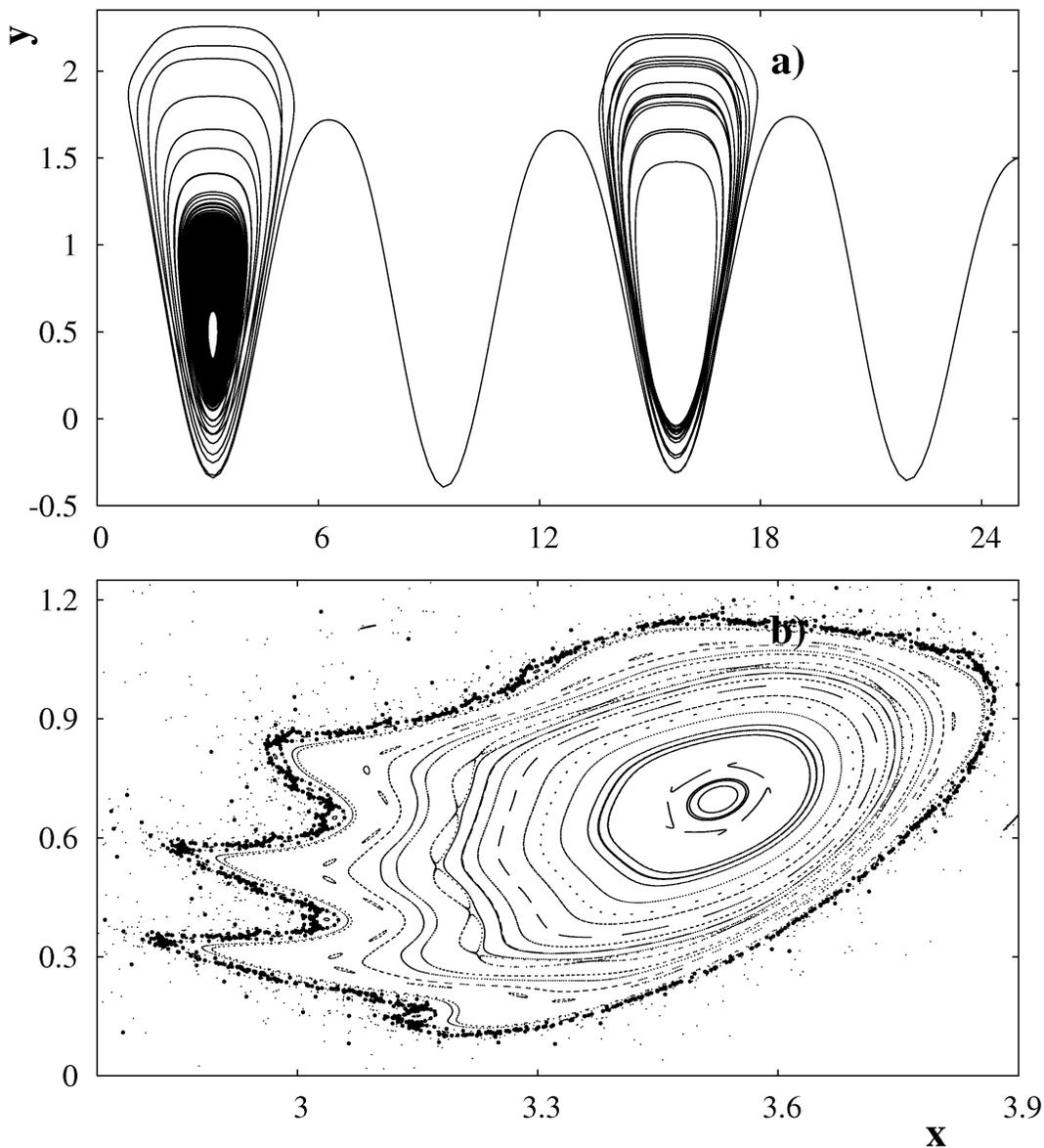}} 
\caption{ 
A single ballistic trajectory (a) sticking to the vortex core (b).
Bold dots show particle's positions through the period of the perturbation. } 
\label{fig4} 
\end{figure} 

Changing the strength $\varepsilon$ and frequency $\omega$ of
the perturbation, we can change the number of overlapping resonances
between the perturbation frequency and the frequencies $f$
of particle's rotations in the circulation zones.
In Fig.~\ref{fig2} we plot a frequency map $f(x_0,y_0)$ that shows by color
the values of $f$ for particles with initial positions
($x_0,y_0$) in the unperturbed system.
\begin{figure}[!htb]                  
\centerline{\includegraphics[width=0.8\textwidth,clip]{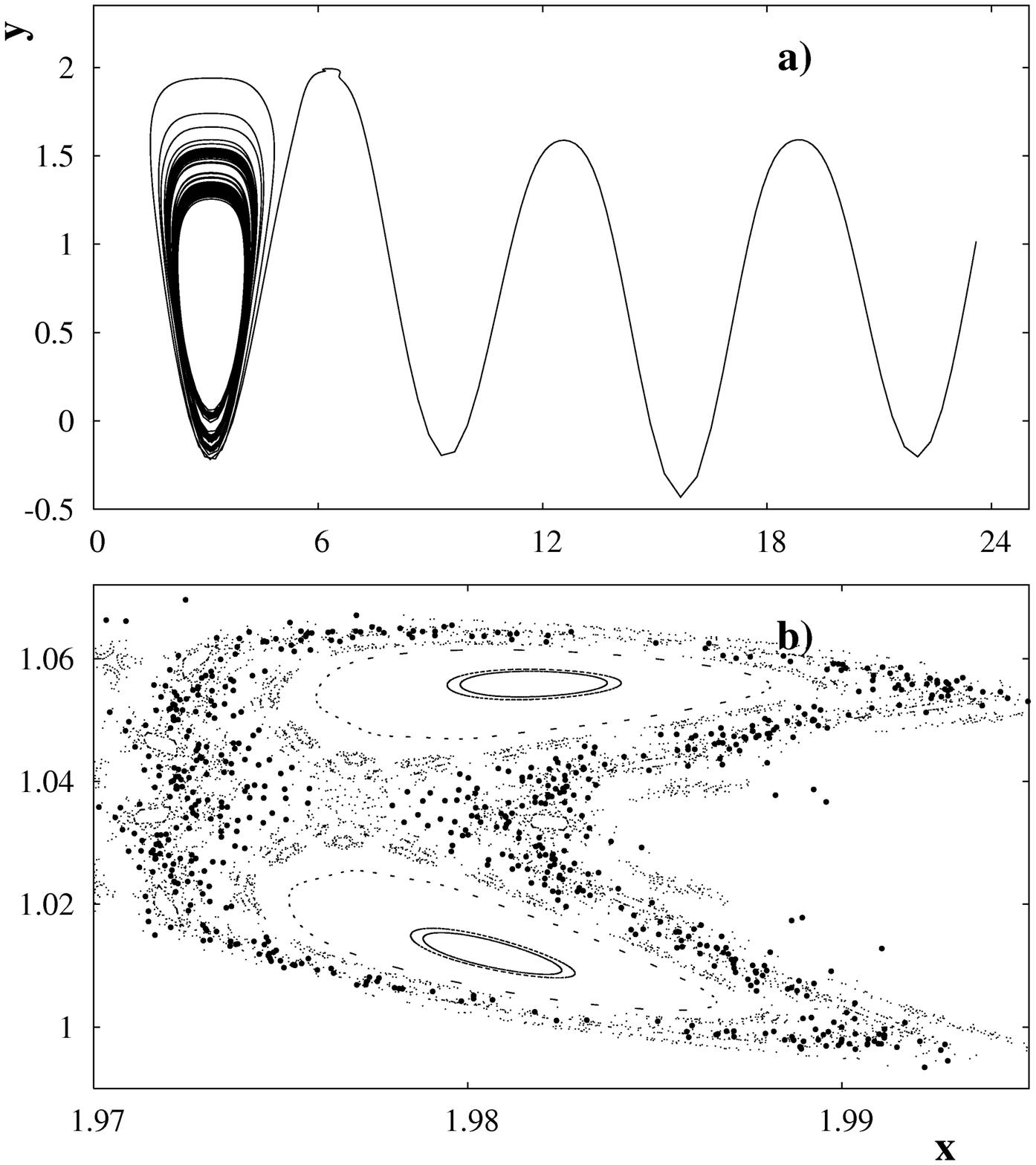}} 
\caption{ 
A single ballistic trajectory (a) sticking to two of the resonant islands of the
secondary resonance (b) Bold dots show particle's positions through the period of the perturbation.} 
\label{fig5} 
\end{figure} 

\begin{figure}[!htb] 
\centerline{\includegraphics[width=0.8\textwidth,clip]{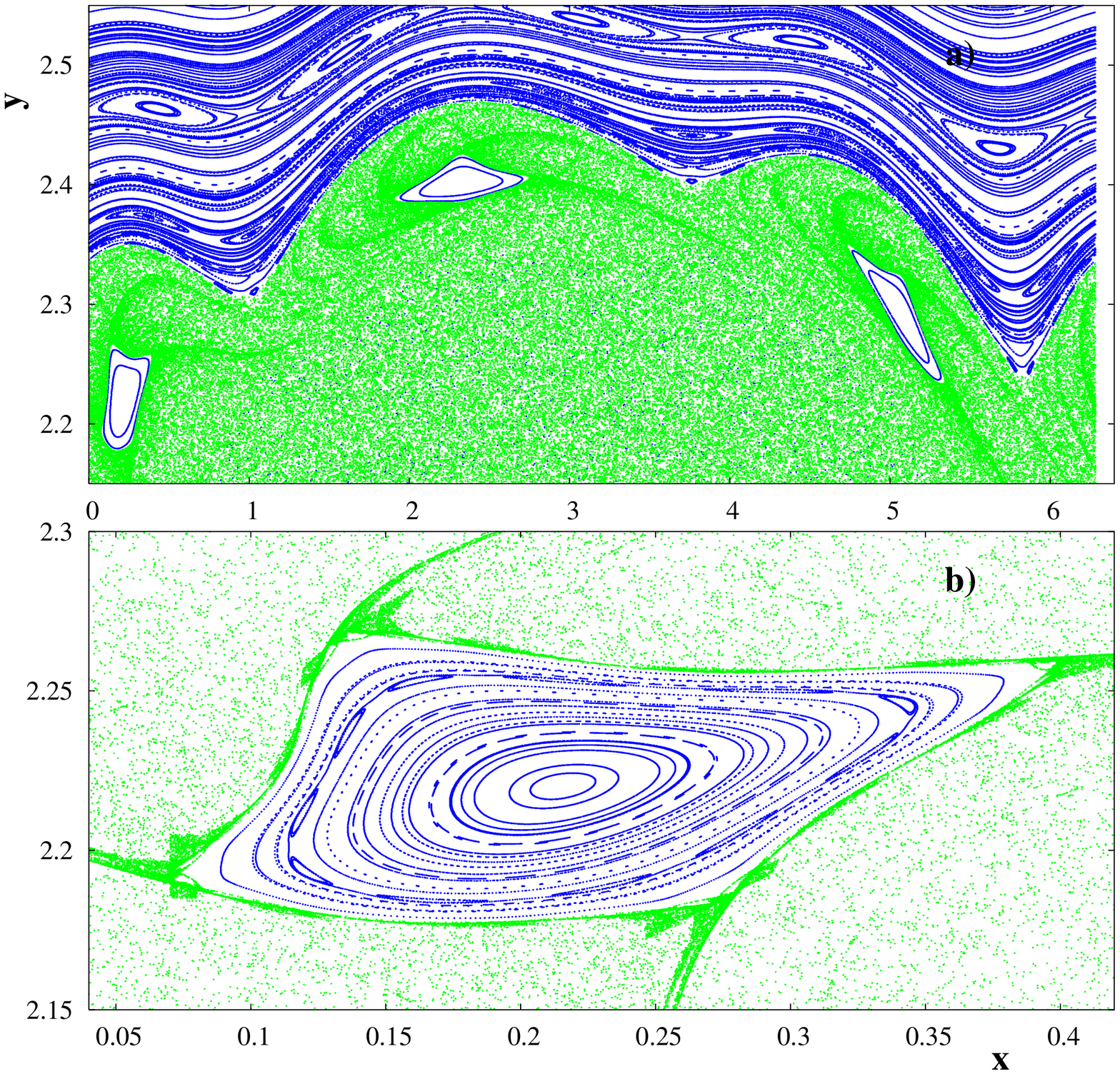}} 
\caption{(color online).  
(a) Chains of ballistic islands on both
sides of the border between the peripheral current and the circulation zone.
(b) Zoom of one of the ballistic islands and its sticky border. } 
\label{fig6} 
\end{figure} 
Let us take the values of the perturbation frequency $\omega=0.2536$ to be close 
to the values of the unperturbed frequency $f$ in the inner core of the circulation 
zone and the strength of
perturbation to be rather small $\varepsilon=0.0785$. Due to the
zonal and meridian symmetries it is possible
to consider particle's motion on the cylinder with $0\le x\le 2\pi$.
The respective Poincar{\'e} section for a large number of trajectories
is shown in Fig.~\ref{fig3}. The vortex core, that survives under this small
perturbation, is immersed into a stochastic sea where one can see six
islands of a secondary resonance to be emerged from
three islands of the primary resonance $3f=2\omega$, where $f=0.169$.
A pair of the secondary resonance islands is zoomed in Fig.~\ref{fig3}b.
\begin{figure}[!htb] 
\centerline{\includegraphics[width=0.8\textwidth,clip]{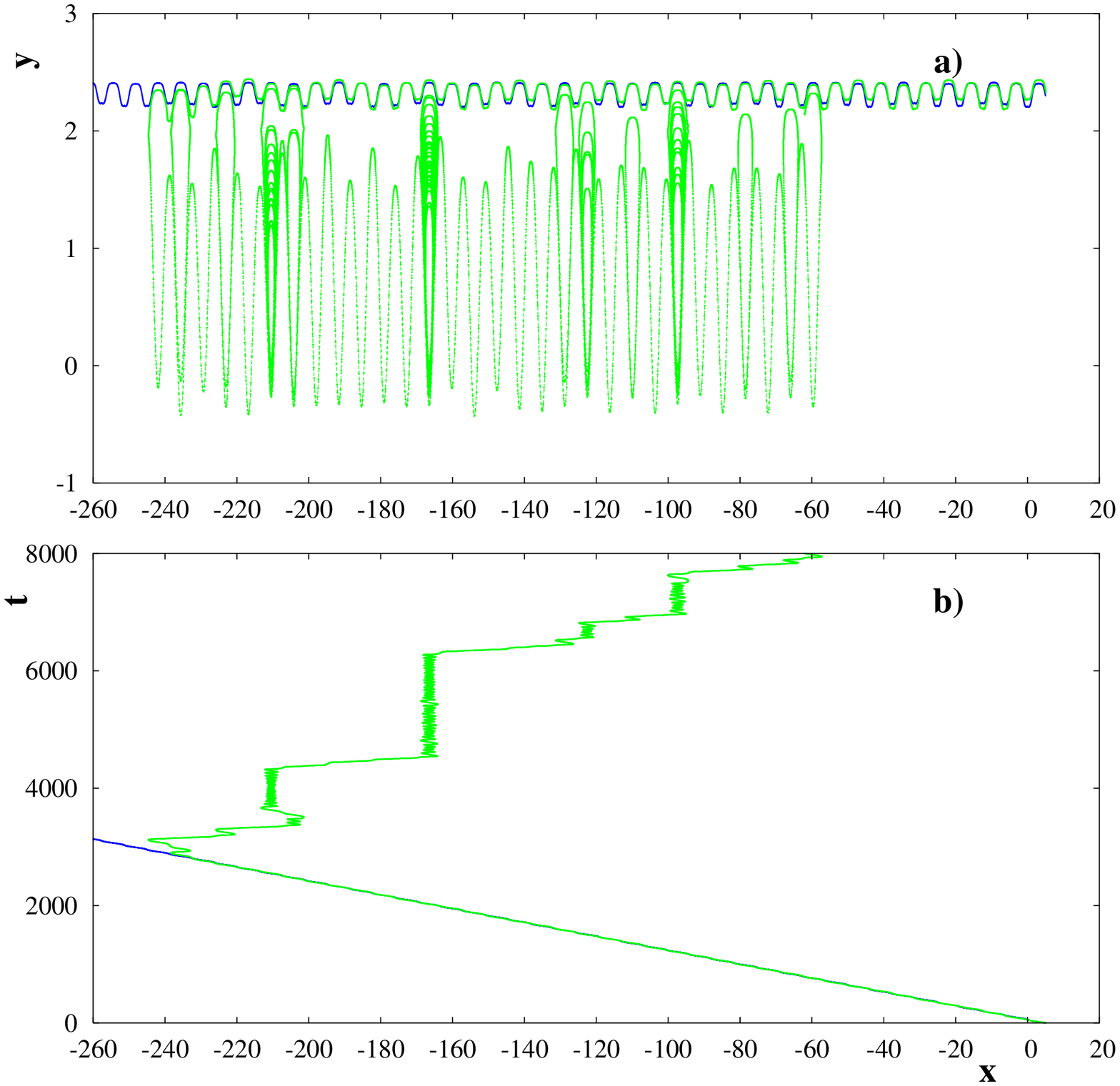}} 
\caption{ (color online). 
Examples of ballistic trajectories (a) on the ($x,y$) plane
and (b) on the ($t,x$) plane. The blue regular trajectory, which is the upper
one in (a) and the lower one in (b), is inside a ballistic island.
The green chaotic trajectory with close initial position, which
is the lower one in (a) and the upper one in (b), demonstrates 
intermittent sticking and flight events.} 
\label{fig7} 
\end{figure} 

Sticking of particles to boundaries between regular and chaotic regions in
the phase space is a typical phenomena with chaotic Hamiltonian
systems \cite{Z02,K83,CS84,M92,BZ93,Z05}. What is special in chaotic advection,
that real fluid parcels may be trapped for a long time near a vortex
core and borders of resonant islands. As to the vortex core, we illustrate
this phenomenon in Fig.~\ref{fig4}, where tracks of a single ballistic 
trajectory (Fig.~\ref{fig4}a) through the period of the perturbation
$2\pi/\omega$ are shown by bold dots (Fig.~\ref{fig4}b).
In Fig.~\ref{fig5} we plot tracks of another ballistic trajectory (Fig.~\ref{fig5}a)
sticking to the islands of the secondary resonance.

Besides the resonant islands with particles moving around the 
elliptic point in the same frame (Fig.~\ref{fig3}), we have found
so-called ballistic islands situated in the chaotic sea, in the peripheral currents and 
on the border between the chaotic sea and the meandering jet. Ballistic modes \cite{Z05,VRKZ99,IGF98} 
correspond to the stable periodic motion of particles
from one frame to another. When mapping their positions
at the moments of time $t_k=2k\pi/\omega$ (k=1,2,\dots) onto the first
frame, one can see chains of ballistic islands along the stochastic layer that replaces the destroyed separatrix (Fig.~\ref{fig6}a).
Zoom of one of the ballistic islands and stickiness of a chaotic
trajectory to its border are shown in Fig.~\ref{fig6}b.
In Fig.~\ref{fig7} we demonstrate two different ballistic trajectories.
If a particle is placed initially inside the ballistic island, it
travels (to the west in this case) in a regular way
(see the upper trajectory in Fig.~\ref{fig7}a) with practically
a constant or slightly modulated zonal velocity (see the lower line in Fig.~\ref{fig7}b).
The lower trajectory in Fig.~\ref{fig7}a on the $(x,y)$ plane
corresponds to a particle placed initially nearby the border of the same
ballistic island from the outside. Dependence of its zonal position 
on time (the upper curve in Fig.~\ref{fig7}b) demonstrates clearly an intermittency of flight
and sticking events. Stickiness to the islands in Figs.~\ref{fig4}~-- ~\ref{fig6} 
is shown for different trajectories, for a convenience. Actually, any chaotic trajectory 
sticks to  different islands at different time intervals. Such a behavior one can see in  
Fig.~\ref{fig7}. This type of dynamics complicates its description because of a multifractal nature. 

\section{\label{Transport}Chaotic transport} 
With increasing the perturbation strength $\varepsilon$, the boundaries
(see Fig.~\ref{fig1}) between the jet ($J$), circulations ($C$), and
peripheral currents ($P$) begin to break down. The vortex core
shrinks (see Fig.~\ref{fig3}), oscillatory and ballistic islands appear
(see Figs.~\ref{fig3} and \ref{fig6}), and a stochastic layer
arises instead of the impenetrable transport barrier, the unperturbed
separatrix. Passive particles which stick to the boundaries of regular
motion can spend there a significant time and deposit to
subdiffusion. Those which find themselves in the jet may travel
long distances to the east between sticking events, whereas the
particles in the peripheral currents travel in the west directions.
Because of the absence of the former transport barriers, the motion of
some particles is intermittent with a large number of turning points
at which particles change directions.
\begin{figure}[!htb] 
\centerline{\includegraphics[width=0.8\textwidth,clip]{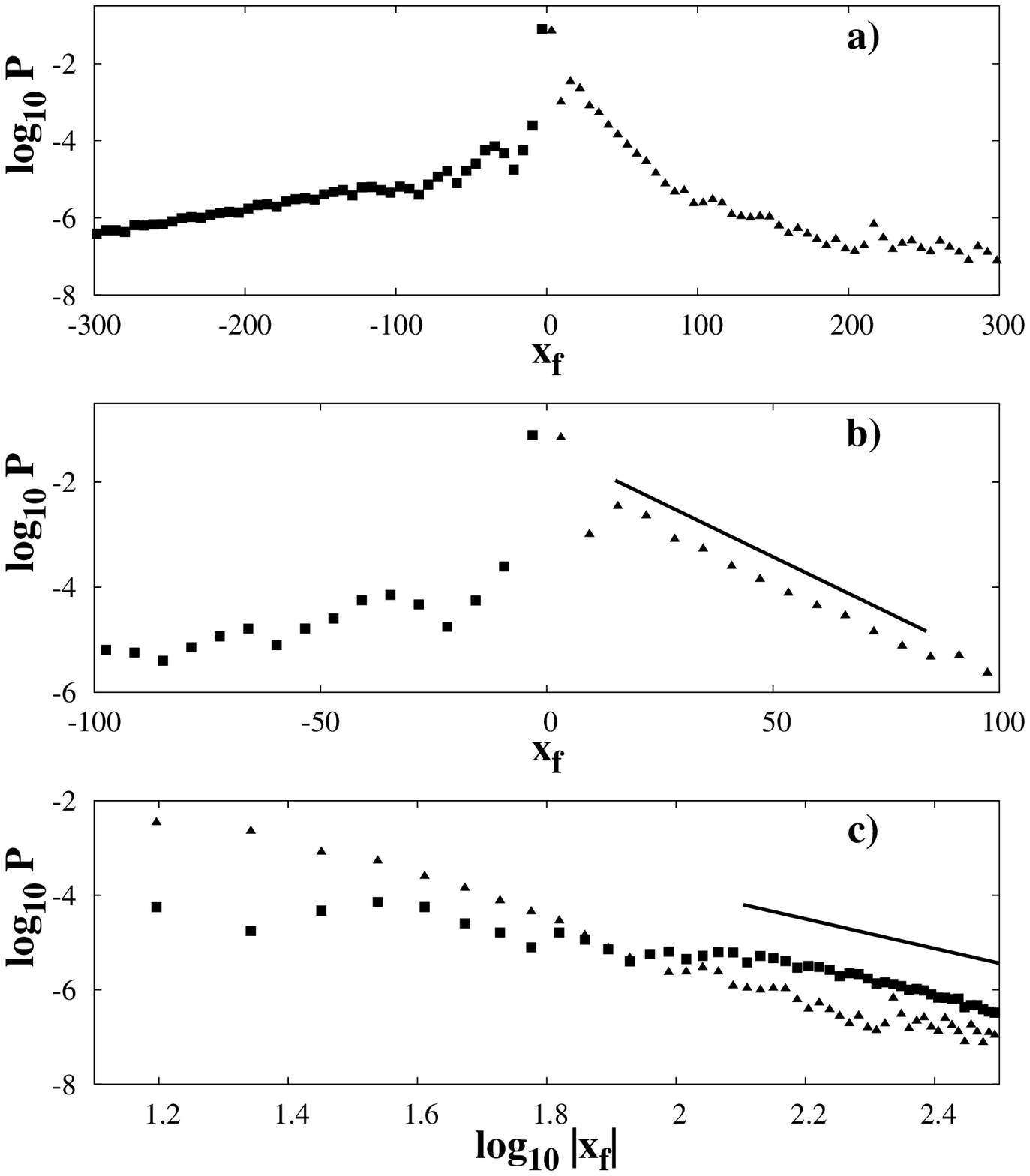}} 
\caption{ 
(a) Probability distribution functions (PDFs) of lengths
of eastward (triangles) and westward (squares) flights
obtained with $6$ chaotic trajectories for $5\cdot 10^8$ steps.
(b) For short eastward flights, $15\le x_f\le 85$, the PDF decays
exponentially, whereas the PDF for westward flights oscillates in the same range. 
(c) The PDF tail of lengths of westward (squares) long flights decays as a power law 
with the slope $\nu=3.12 \pm 0.1$. 
} 
\label{fig8} 
\end{figure} 
\begin{figure}[!htb] 
\centerline{\includegraphics[width=0.8\textwidth,clip]{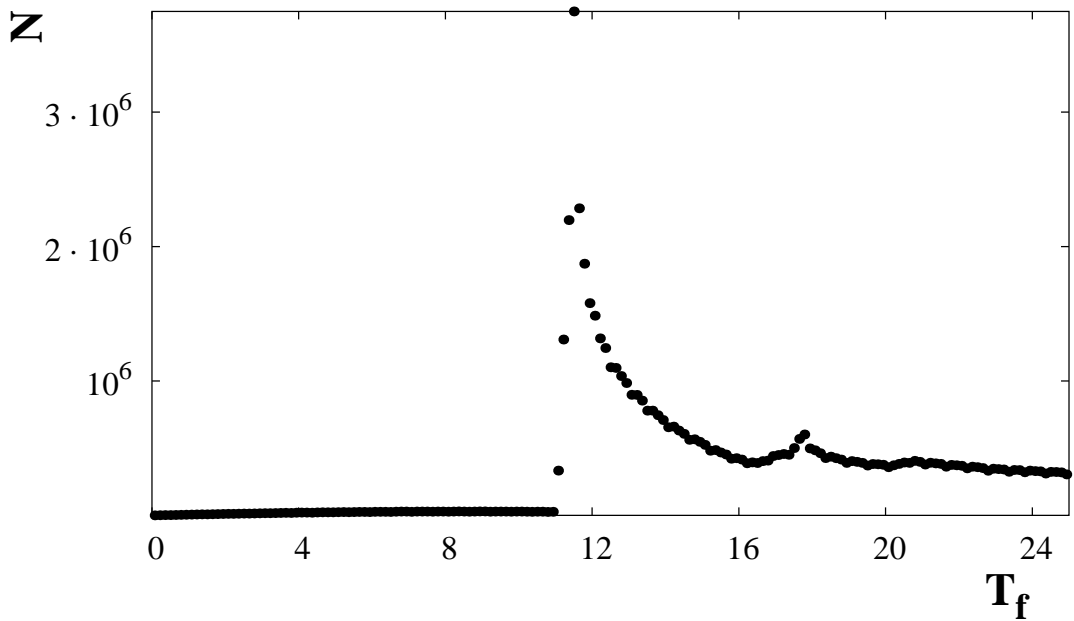}} 
\caption{ 
The number of particles $N$ executing short flights with $|x_f|<2\pi$ 
and a fixed duration $T_f$. Peaks at $T_f\simeq 11.8$
and $T_f\simeq 18$ correspond to particles sticking
to the boundaries of the vortex core and oscillatory islands.} 
\label{fig9} 
\end{figure} 
\begin{figure}[!htb] 
\centerline{\includegraphics[width=0.8\textwidth,clip]{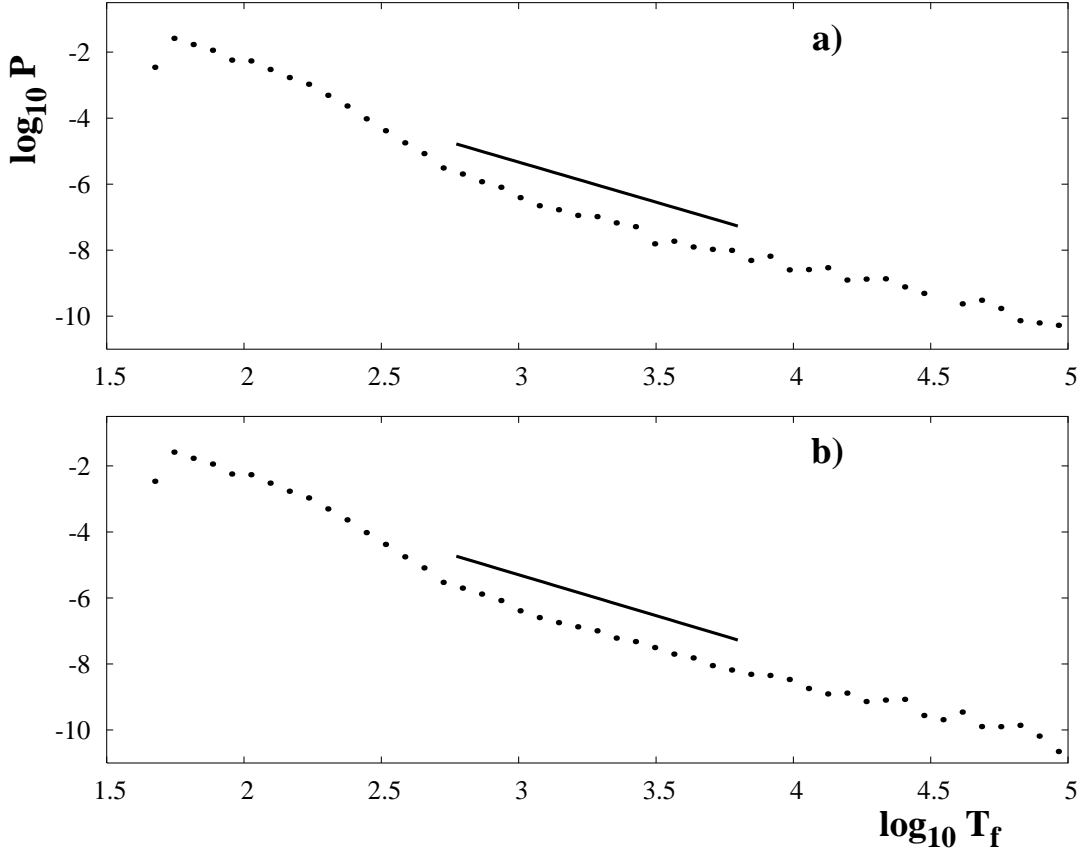}} 
\caption{ 
The PDF of durations of eastward flights (a) with $125$ chaotic trajectories for $5\cdot 10^6$  steps (the slope is $\gamma=2.43 \pm 0.1$) and (b) with $6$ chaotic trajectories for $5\cdot 10^8$  steps (the slope is $\gamma=2.48 \pm 0.05$).
} 
\label{fig10} 
\end{figure} 
\begin{figure}[!htb] 
\centerline{\includegraphics[width=0.8\textwidth,clip]{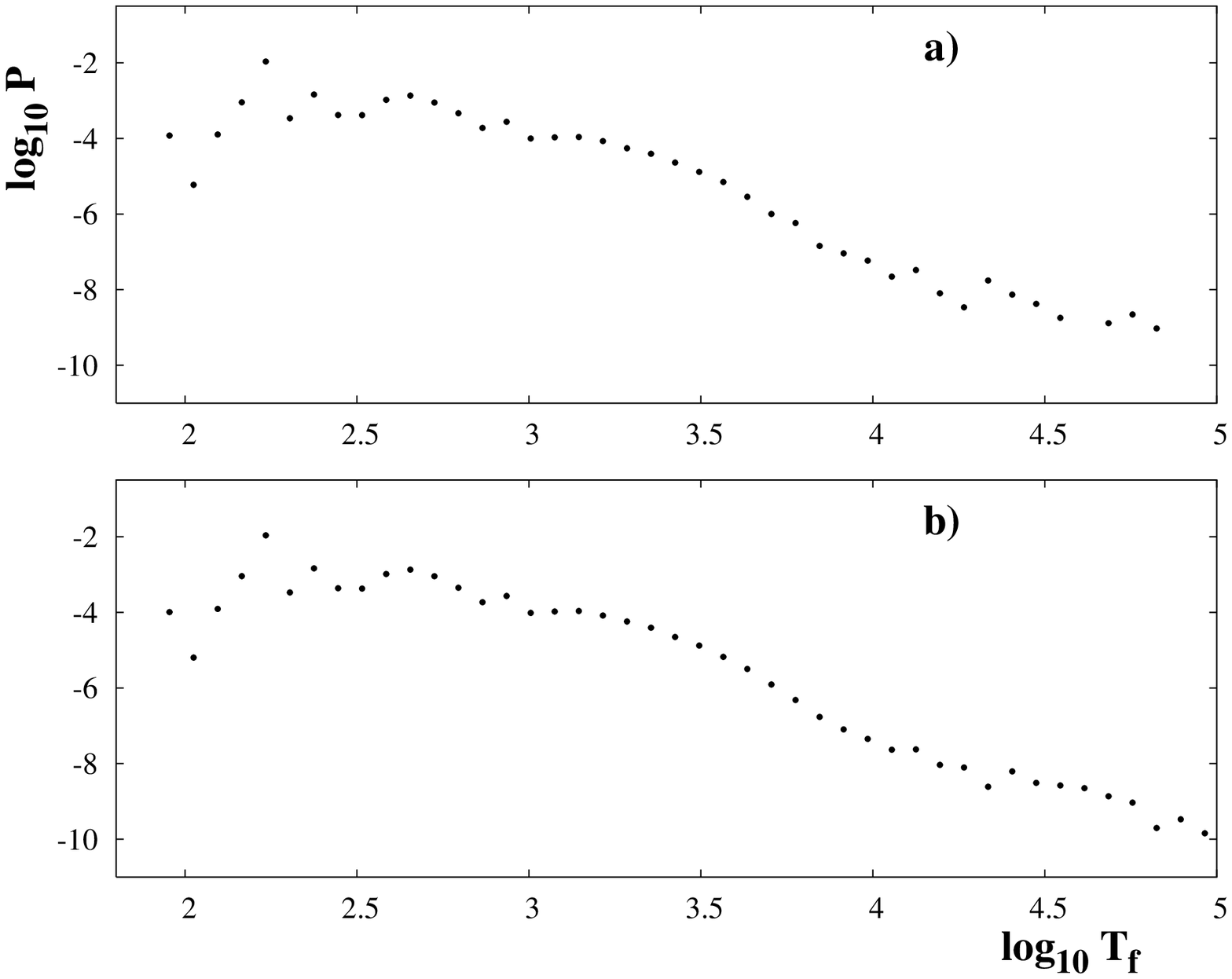}} 
\caption{ 
The  PDF of durations of westward flights (a) with $125$ chaotic trajectories for $5\cdot 10^6$  steps  and (b) with $6$ chaotic trajectories for $5\cdot 10^8$  steps.} 
\label{fig11} 
\end{figure} 
As usual, we will characterize statistical properties of particles motion
by probability distribution functions (PDFs). We will
call ``a flight'' any event between two successive changes of signs
of the particle's zonal velocity. In this terminology
a sticking consists of a number of flights with approximately equal
flight times. It should be noted that some authors distinguish
flights from sticking motion by examining a distance between successive local
extrema of trajectories $x(t)$ \cite{Ko00, SWS94}.
Both the PDF of durations of flights $P(T_f)$ and of lengths of flights 
$P(x_f)$ will be used to characterize the chaotic transport. 
\begin{figure}[!htb] 
\centerline{\includegraphics[width=0.8\textwidth,clip]{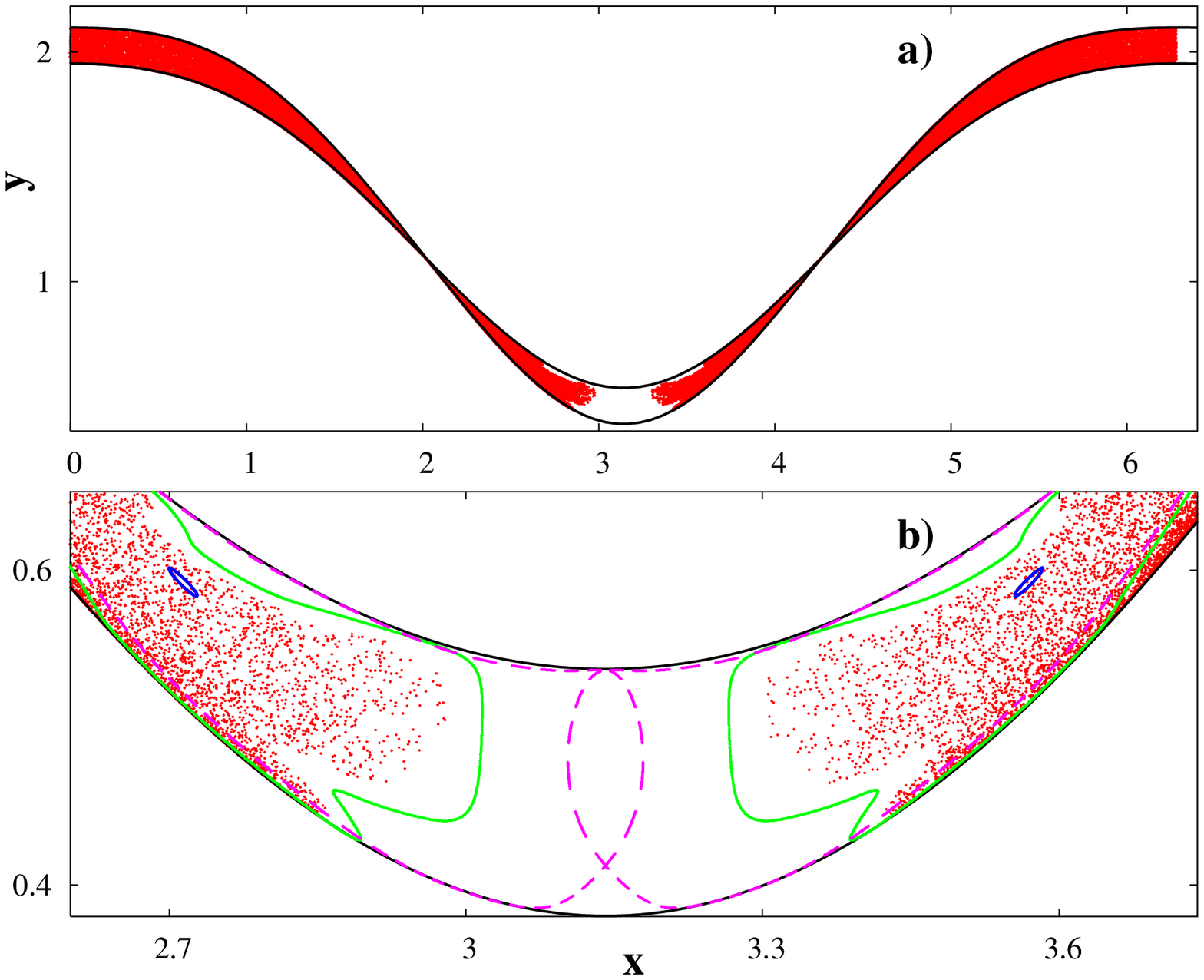}} 
\caption{ (color online). 
Turning points (red ones online) of a single chaotic trajectory
on the cylinder $0\le x\le2\pi$ are in a narrow strip confined 
by two curves with equations $y= f(x)$ to be defined in the text.
(a) General view. (b) Zoom with a few regular trajectories shown for reference.} 
\label{fig12} 
\end{figure} 

The Poincar{\'e} section of the flow with the perturbation
strength $\varepsilon=0.0785$ and frequency $\omega=0.2536$ is
shown in Fig.~\ref{fig3}a. The flight PDFs are computed with 
$6$ particles (initially placed in the first east frame inside
a stochastic layer) up to the long time $t=5\cdot 10^8$ and 
with $125$ particles up to $t=5\cdot 10^6$. Particles inside 
the stochastic layer may execute very complicated motion changing the direction in 
a chaotic way, sticking to the island's boundaries and executing ballistic flights 
with different values of length and duration. 

The PDF of the lengths of flights is shown in Fig.~\ref{fig8}a for both the
directions. The asymmetry between the eastward (shown by triangles)
and westward (shown by squares) flights is evident. The transport velocity 
in the peripheral currents is different from  the transport velocity in the jet. 
Mixing between the circulations and the peripheral currents is, in general, different 
from mixing between the circulations and the jet. 
Both the PDFs can be roughly split into three distinctive regions. The very short
flights with small values of $|x_f|$ ($<2\pi$) are supposed
to be dominated by sticking to the boundaries of the vortex core and
oscillatory islands. To check that we plot in Fig.~\ref{fig9} the number 
of particles $N$ executing short flights with small values of $T_f$ and with the 
lengths of flights $|x_f|<2\pi$.
The prominent peaks at $T_f\simeq 11.8$ and $T_f\simeq 18$ correspond
to those particles which  stick to the vortex core and resonant island's boundaries
(see Fig.~\ref{fig3} and ~\ref{fig4}). 
The PDF for eastward flights with the lengths in the range
$15\le x_f\le 85$ decays exponentially, whereas  the PDF for westward flights 
is an oscillating function in this range (see Fig.~\ref{fig8}b).
The tails of both the PDFs are close to a power-law decay
$P_f(x)\sim |x|^{-\nu}$. In Fig.~\ref{fig8}c we estimate
the exponent for long westward flights with corresponding
error by least-square fitting of the straight line 
to the log-log plot of the data to be $\nu=3.12 \pm 0.1$. The tail for  
long eastward flights in the same range is an oscillating function whose slope is difficult to 
measure because of insufficient statistics. Simulation with 125 initial conditions up to the time 
$t=5\cdot 10^6$ shows practically the same results.  

The PDF of durations $T_f$ of eastward  flights is shown in Fig.~\ref{fig10} 
with different number of trajectories to be computed and up to different times. In Fig.~\ref{fig10}a we plot the time PDF with $125$ chaotic trajectories for $5\cdot 10^6$  steps. The exponent of the function $P_f(t)\sim t^{-\gamma}$ is estimated to be $\gamma=2.43 \pm 0.1$ in the range from $\log_{10}T_f = 2.75$ to $\log_{10}T_f = 3.8$. In Fig.~\ref{fig10}b  the time PDF is plotted with $6$ chaotic trajectories but for much longer times $5\cdot 10^8$. The respective exponent is estimated to be practically the same, $\gamma=2.48 \pm 0.05$. 
What is remarkable is that both the time PDFs demonstrate at the very 
tails periodic oscillations in the logarithmic scale. That is a 
common feature with time distribution functions in typical Hamiltonian systems 
\cite{Z02,Z05,Ko00}. One of the possible 
explanations of this fact is existence of a fractal-like hierarchy of 
islands  \cite{Z02} and a discrete renormalization group \cite{Sor98}. 
The respective PDF of 
durations of westward  flights is shown in Fig.~\ref{fig11} with 
$125$ chaotic trajectories for $5\cdot 10^6$  steps 
(Fig.~\ref{fig11}a) and  with $6$ chaotic trajectories for $5\cdot 10^8$  
steps 
(Fig.~\ref{fig11}b). The single algebraic tail is not so evident with  
westward  flights   as with eastern ones. Rather one can see in this 
case more prominent log-periodic oscillations. 

As it is seen from Fig.~\ref{fig10}, the flight-time distribution  
$P_f(t)$ has the asymptotic 
\begin{equation}
P_f(t) = \frac{{\rm const}}{t^{\gamma}}. 
\label{flight}
\end{equation}
One can consider it as an intermediate asymptotic  depending on which set of sticky islands it represents. 
We can expect a fairly complicate multifractal dependence of $P_f$ on $t$ and the simplified formula 
(\ref{flight}) should be applied to specific interval of time and parameters (see more discussion in 
Refs. \cite{Z02,Z05}). Another approach to study the stickiness is to consider a distribution of Poincar{\'e}  
recurrences of trajectories $P_{\rm rec}(t,B)$ to a small domain $B$ taken in a stochastic sea far enough 
from sticky domains. It was shown in Refs. \cite{Z02,Z05} that for many different models 
of chaotic Hamiltonian dynamics the stickiness leads to a distribution of density probability to find 
the recurrence time within the interval $(t,t+dt)$: 
\begin{equation}
P_{\rm rec}(t,B) = \frac{{\rm const}}{t^{\gamma^{\rm rec}}}, 
\label{rec}
\end{equation}
where $\gamma^{\rm rec}= \min \gamma$, $(t \to \infty)$. From Fig.~\ref{fig10} we have 
$\gamma= 2.43 \pm 0.1)$ and $\gamma= 2.48 \pm 0.05)$. This result is consistent with the Kac lemma \cite{Z02,Z05} that imposes $\gamma >2$. 

A typical chaotic particle in a stochastic
layer changes many times its direction (the sign of the zonal velocity $u$).
We have found that the respective turning points
are in a narrow strip confined by the curves to be defined by 
the unperturbed equations of motion (\ref{main_sys})
with $\dot x=0$ and $A=A_0 \pm \varepsilon$.
As an example, we demonstrate in Fig.~\ref{fig12}a (color online) a narrow
strip of red turning points confined by the curves with
$A_0=0.785$ and $\varepsilon=0.0785$. All these points belong
to a single chaotic trajectory on the cylinder $0\le x\le 2\pi$.
Figure~\ref{fig12}b provides a zoom of the lower part
of the strip with the confining curves
\begin{equation}
%\begin{multline}
y=L\sqrt{1+A^2\sin^2{x}}\operatorname{Arcosh}{\sqrt{\frac{1}{LC
\sqrt{1+A^2\sin^2{x}}}}}+A\cos x
%\end{multline}
\label{strip}
\end{equation}

and $A=A_0 \pm \varepsilon$ to be shown by black solid lines and three regular
trajectories to be shown by different colors. The empty part of the strip
corresponds to the place where the vortex core is situated
(see Fig.~\ref{fig3}a).

\section{\label{SecConc}Conclusion} 

In this paper we have investigated mixing and transport of passive particles in a two-dimensional 
nonstationary shear flow of the ideal fluid that is relevant to study of mixing of water 
masses along with their properties in jet-like western boundary currents in the oceans.

We have considered a known kinematic model of a meandering $\operatorname{sech}^2$ zonal flow \cite{B89,S92}, 
derived the respective advection equations in the reference frame, moving with the phase velocity of the meander, 
found the stationary points and their stability, and possible bifurcations. The control parameters of 
the model streamfunction (\ref{psi}), the scaled jet's width $L=0.628$, the phase velocity $C=0.1168$, and the meander's amplitude $A=0.785$, were chosen to correspond to estimated realistic values of the Gulf Stream meander's parameters 
\cite{B89,S92} and to the topology of the flow with two chains of circulation cells separated by  an eastward jet 
(the case 2b in the list of possible flow regimes in Sec.~\ref{Model}). 

The respective phase space has been shown to consist of the vortex cores filled with regular trajectories surrounded by a chaotic sea 
with stochastic layers and chains of islands of oscillatory and ballistic regular motion. The  boundaries of these regular structures have been shown to be sticky. Typical advected particles may alternate chaotically between being trapped 
near the vortex cores and island's boundaries and ``flying'' freely in the jet. 

Defining ``a flight'' as an event between two successive changes of signs
of the particle's zonal velocity, we characterize the statistics of flights by the PDFs computing both the PDF  of durations of the flights $P_f(t)$ and of lengths of the flights $P_f(x)$. 
Both the   functions have been found to have regions with different laws. For short flights we have found two  peaks in the $P_f(x)$ function 
corresponding to sticking events in the circulations, for middle flights an exponential decay has been found, and the tails with long flights has been found to fit to a  power law. An asymmetry of transport of passive particles in the eastern and western directions is caused by a difference in  mixing between the circulations and the peripheral currents and  mixing between the circulations and the jet.
The tails of the time PDFs for eastern and western flights 
have been shown to be close to  power-law decay functions. The anomalous statistics of the flights is caused by the coherent structures -- the circulation cells, the peripheral currents, and the jet -- where particles can spend an anomalous amount of time either being trapped in the circulations or moving in the peripheral currents and in the jet.

\section*{Acknowledgments} 
S.P., M.B., and M.U. were supported by the Russian Foundation for Basic Research 
(Grant  no.06-05-96032), by the Program ``Mathematical Methods
in  Nonlinear Dynamics'' of the Russian Academy of Sciences  
and by the Program for Basic Research of the Far Eastern Division of the Russian Academy of
Sciences. G.Z. and S.P. were supported by the ONR Grant no. N00014-02-1-0056 and the NSF Grant no. DMS-0417800.

\end{document}